\documentclass[epj]{webofc}
\usepackage[utf8]{inputenc}
\usepackage[varg]{txfonts}   
\usepackage{booktabs}
\usepackage{xcolor}
\definecolor{darkred}{rgb}{0.4,0.0,0.0}
\definecolor{darkgreen}{rgb}{0.0,0.4,0.0}
\definecolor{darkblue}{rgb}{0.0,0.0,0.4}
\usepackage[bookmarks,linktocpage,colorlinks,
    linkcolor = darkred,
    urlcolor  = darkblue,
    citecolor = darkgreen]{hyperref}
\usepackage{subfigure}
\usepackage{amsmath,amssymb}
\usepackage{dsfont}
\usepackage{slashed}
\usepackage{verbatim}    
\usepackage{graphicx} 

\newcommand{\be}{\begin{equation}}
\newcommand{\ee}{\end{equation}}
\newcommand{\bea}{\begin{eqnarray}} 
\newcommand{\eea}{\end{eqnarray}}

\newcommand{\MSbar}{{\overline{\rm MS}}}

\newcommand{\gtilde}{\frac{g^2}{16 \, \pi^2}\; }
\newcommand{\qslash}{{\not{\hspace{-0.07cm}q}}}

\newcommand{\la}{\lambda}

\wocname{EPJ Web of Conferences}
\woctitle{Lattice2017}
\def\openone{\leavevmode\hbox{\small1\kern-3.3pt\normalsize1}}

\begin{document}
\selectlanguage{english}
\title{Renormalization of Supersymmetric QCD on the Lattice}

\author{\firstname{Marios} \lastname{Costa}\inst{1}\fnsep\thanks{Speaker, \email{kosta.marios@ucy.ac.cy}}  \and
\firstname{Haralambos} \lastname{Panagopoulos}\inst{1} }

\institute{Department of Physics, University of Cyprus, CY-1678 Nicosia, Cyprus}

\abstract{We perform a pilot study of the perturbative renormalization of a Supersymmetric gauge theory with matter fields on the lattice. As a specific example, we consider Supersymmetric ${\cal N}{=}1$ QCD (SQCD). We study the self-energies of all particles which appear in this theory, as well as the renormalization of the coupling constant. To this end we compute, perturbatively to one-loop, the relevant two-point and three-point Green's functions using both dimensional and lattice regularizations. Our lattice formulation involves the Wilson discretization for the gluino and quark fields; for gluons we employ the Wilson gauge action; for scalar fields (squarks) we use naive discretization. The gauge group that we consider is $SU(N_c)$, while the number of colors, $N_c$, the number of flavors, $N_f$, and the gauge parameter, $\alpha$, are left unspecified.

We obtain analytic expressions for the renormalization factors of the coupling constant ($Z_g$) and of the quark ($Z_\psi$), gluon ($Z_u$), gluino ($Z_\lambda$), squark ($Z_{A_\pm}$), and ghost ($Z_c$) fields on the lattice. We also compute the critical values of the gluino, quark and squark masses. Finally, we address the mixing which occurs among squark degrees of freedom beyond tree level: we calculate the corresponding mixing matrix which is necessary in order to disentangle the components of the squark field via an additional finite renormalization. 
}

\maketitle

\section{Introduction}\label{intro}
The current intensive searches for Physics Beyond the Standard Model (BSM) are becoming a very timely endeavor, given the precision experiments at LHC and elsewhere; at the same time, numerical studies of BSM Physics are more viable due to the advent of lattice formulations which preserve chiral symmetry. Furthermore, the lattice formulation of various supersymmetric models is currently under active study ~\cite{josephREVIEW, Catterall:2009it}. As a forerunner to a long-term prospect of addressing numerically supersymmetric extensions of the Standard Model, we have undertaken an investigation of SQCD, in order to address some of the fundamental difficulties which must be resolved before further progress can be made. Within the SQCD formulation we compute the quark ($\psi$), gluino ($\lambda^\alpha$), gluon ($u_\mu^\alpha$), squark ($A$) propagators and the gluon-antighost-ghost Green's function. Our computations are performed to one loop and to lowest order in the lattice spacing, $a$. We extract the renormalization factors for the coupling constant ($Z_g$) and of the quark ($Z_\psi$), gluon ($Z_u$), gluino ($Z_\lambda$), squark ($Z_{A_\pm}$), and ghost ($Z_c$) fields and the critical masses of quark, gluino and squark fields.

The details of our work, along with a longer list of references, can be found in Ref.\cite{Costa:2017rht}.

\section{Lattice Action}\label{latticeS}
Even though the lattice breaks supersymmetry explicitly \cite{Curci&Venz}, it is the only regulator which describes many aspects of strong interactions nonperturbatively. We will extend Wilson's formulation of the QCD action, to encompass SUSY partner fields as well. In this standard discretization, quarks, squarks and gluinos live on the lattice sites, and gluons live on the links of the lattice: $U_\mu (x) = e^{i g a T^{\alpha} u_\mu^\alpha (x+a\hat{\mu}/2)}$. This formulation leaves no SUSY generators intact, and it also breaks chiral symmetry; it thus represents a ``worst case'' scenario, which is worth investigating in order to address the complications \cite{Giedt} which will arise in numerical simulations of SUSY theories. In our ongoing investigation we plan to address also improved actions, so that we can check to what extent some of the SUSY breaking effects can be alleviated.

For Wilson-type quarks ($\psi$) and gluinos ($\lambda$), the Euclidean action ${\cal S}^{L}_{\rm SQCD}$ on the lattice becomes ($A_\pm$: squark field components):       
\bea
{\cal S}^{L}_{\rm SQCD} & = & a^4 \sum_x \Big[ \frac{N_c}{g^2} \sum_{\mu,\,\nu}\left(1-\frac{1}{N_c}\, {\rm Tr} U_{\mu\,\nu} \right ) + \sum_{\mu} {\rm Tr} \left(\bar \lambda_M \gamma_\mu {\cal{D}}_\mu\lambda_M \right ) - a \frac{r}{2} {\rm Tr}\left(\bar \lambda_M  {\cal{D}}^2 \lambda_M \right) \nonumber \\ 
&+&\sum_{\mu}\left( {\cal{D}}_\mu A_+^{\dagger}{\cal{D}}_\mu A_+ + {\cal{D}}_\mu A_- {\cal{D}}_\mu A_-^{\dagger}+ \bar \psi_D \gamma_\mu {\cal{D}}_\mu \psi_D \right) - a \frac{r}{2} \bar \psi_D  {\cal{D}}^2 \psi_D \nonumber \\
&+&i \sqrt2 g \big( A^{\dagger}_+ \bar{\lambda}^{\alpha}_M T^{\alpha} P_+ \psi_D  -  \bar{\psi}_D P_- \lambda^{\alpha}_M  T^{\alpha} A_+ +  A_- \bar{\lambda}^{\alpha}_M T^{\alpha} P_- \psi_D  -  \bar{\psi}_D P_+ \lambda^{\alpha}_M  T^{\alpha} A_-^{\dagger}\big)\nonumber\\  
&+& \frac{1}{2} g^2 (A^{\dagger}_+ T^{\alpha} A_+ -  A_- T^{\alpha} A^{\dagger}_-)^2 - m ( \bar \psi_D \psi_D - m A^{\dagger}_+ A_+  - m A_- A^{\dagger}_-)\Big] \,,
\label{susylagrLattice}
\eea
where: $U_{\mu \nu}(x) =U_\mu(x)U_\nu(x+a\hat\mu)U^\dagger_\mu(x+a\hat\nu)U_\nu^\dagger(x)$, and a summation over flavors is understood in the last three lines of Eq.~(\ref{susylagrLattice}). The 4-vector $x$ is restricted to the values $x = na$, with $n$ being an integer 4-vector. 
Thus the momentum integration, after a Fourier transformation, is restricted to the first Brillouin zone (BZ) $[ - \pi/a,\pi/a ]^4$ and the sum over $x$ leads to momentum conservation in each vertex. 
The terms proportional to the Wilson parameter, $r$, eliminate the problem of fermion doubling, at the expense of breaking chiral invariance.

The definitions of the covariant derivatives are as follows:
\bea
{\cal{D}}_\mu\lambda_M(x) &\equiv& \frac{1}{2a} \Big[ U_\mu (x) \lambda_M (x + a \hat{\mu}) U_\mu^\dagger (x) - U_\mu^\dagger (x - a \hat{\mu}) \lambda_M (x - a \hat{\mu}) U_\mu(x - a \hat{\mu}) \Big] \\
{\cal D}^2 \lambda_M(x) &\equiv& \frac{1}{a^2} \sum_\mu \Big[ U_\mu (x)  \lambda_M (x + a \hat{\mu}) U_\mu^\dagger (x)  - 2 \lambda_M(x) \nonumber\\
&& \hspace{1cm}+  U_\mu^\dagger (x - a \hat{\mu}) \lambda_M (x - a \hat{\mu}) U_\mu(x - a \hat{\mu})\Big]\\
{\cal{D}}_\mu \psi_D(x) &\equiv& \frac{1}{2a}\Big[ U_\mu (x) \psi_D (x + a \hat{\mu})  - U_\mu^\dagger (x - a \hat{\mu}) \psi_D (x - a \hat{\mu})\Big]\\  
{\cal D}^2 \psi_D(x) &\equiv& \frac{1}{a^2} \sum_\mu \Big[U_\mu (x) \psi_D (x + a \hat{\mu})  - 2 \psi_D(x) +  U_\mu^\dagger (x - a \hat{\mu}) \psi_D (x - a \hat{\mu})\Big]\\
{\cal{D}}_\mu A_+(x) &\equiv& \frac{1}{a} \Big[  U_\mu (x) A_+(x + a \hat{\mu}) - A_+(x)   \Big]\\
{\cal{D}}_\mu A_+^{\dagger}(x) &\equiv& \frac{1}{a} \Big[A_+^{\dagger}(x + a \hat{\mu}) U_\mu^{\dagger}(x)  -  A_+^\dagger(x)\Big]\\
{\cal{D}}_\mu A_-(x) &\equiv& \frac{1}{a} \Big[A_-(x + a \hat{\mu}) U_\mu^{\dagger}(x)  -  A_-(x)\Big]\\
{\cal{D}}_\mu A_-^{\dagger}(x) &\equiv& \frac{1}{a} \Big[U_\mu (x) A_-^{\dagger}(x + a \hat{\mu})   -  A_-^{\dagger}(x) \Big]
\eea
A gauge-fixing term, together with the compensating ghost field term, must be added to the action, in order to avoid divergences from the  integration over gauge orbits; these terms are the same as in the non-supersymmetric case. Similarly, a standard ``measure'' term must be added to the action, in order to account for the Jacobian in the change of integration variables: $U_\mu \to u_\mu$\,.

\section{The one-loop Feynman diagrams on the lattice}
\label{calculation}
We calculate perturbatively 2-pt and 3-pt Green's functions up to one loop, both in the continuum and on the lattice. The quantities that we study are the self-energies of the quark ($\psi$), gluon ($u_\mu$), squark ($A$), gluino ($\lambda$), and ghost ($c$) fields, using both dimensional regularization (DR) and lattice regularization (L). In addition we calculate the gluon-antighost-ghost Green's function in order to renormalize the coupling constant ($g$). The Green's functions leading to self-energies of squarks exhibit also mixing among $A_+$ and $A_-^\dagger$; we calculate the elements of the corresponding $2\times2$ mixing matrix.

\begin{figure}[ht!]
\centering
\includegraphics[scale=0.7]{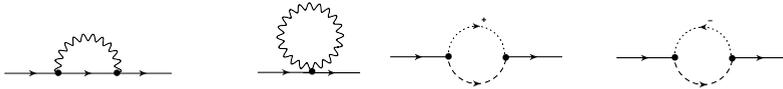}
\caption{One-loop Feynman diagrams contributing to the 2-pt Green's function $\langle \psi(x) \bar \psi(y) \rangle$. A wavy (solid) line represents gluons (quarks). A dotted (dashed) line corresponds to squarks (gluinos). Squark lines are further marked with a $+$($-$) sign, to denote an $A_+ \, (A_-)$ field. A squark line arrow entering (exiting) a vertex denotes a $A_+$ ($A_+^{\dagger}$) field; the opposite is true for $A_-$ ($A_-^{\dagger}$) fields.
  }
\label{quark2pt}
\end{figure}

\begin{figure}[ht!]
\centering
\includegraphics[scale=0.7]{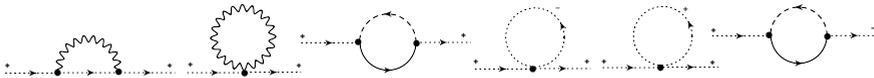}
\caption{The first 5 Feynman diagrams contribute to the 2-pt Green's function 
$\langle A_+(x) A_+^{\dagger}(y) \rangle$. The case of $\langle  A_-^{\dagger}(x) A_-(y) \rangle$ is completely analogous. The last diagram is responsible for mixing between $A_+$ and $A_-^\dagger$.
  }
\label{squark2pt}
\end{figure}

\begin{figure}[h!]
\centering
\includegraphics[scale=0.7]{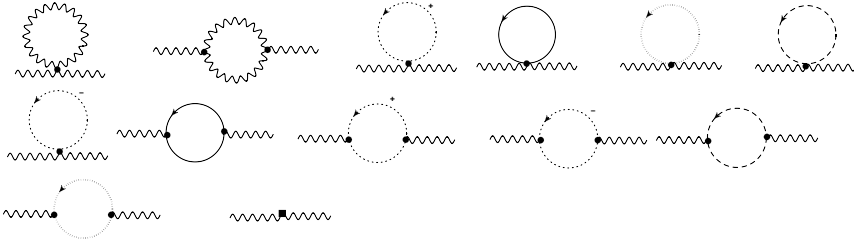}
\caption{One-loop Feynman diagrams contributing to the 2-pt Green's function  $\langle  u_\mu^{\alpha}(x) u_\nu^{\beta}(y) \rangle$. 
The ``double dashed'' line is the ghost field and the solid box in the bottom right vertex comes from the measure part of the lattice action. }
\label{gluon2pt}
\end{figure}

\begin{figure}[h!]
\centering
\includegraphics[scale=0.7]{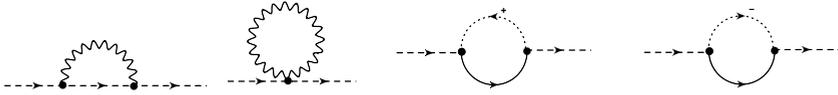}
\caption{One-loop Feynman diagrams contributing to the 2-pt Green's function  $\langle  \lambda^\alpha(x) \bar \lambda^\beta(y) \rangle$.
  }
\label{gluino2pt}
\end{figure}

\begin{figure}[h!]
\centering
\includegraphics[scale=0.7]{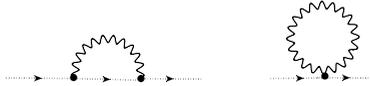}
\caption{One-loop Feynman diagram contributing to the 2-pt Green's function $\langle  c(x) \bar c(y) \rangle$.
  }
\label{ghost2pt}
\end{figure}

\begin{figure}[h!]
\centering
\includegraphics[scale=0.7]{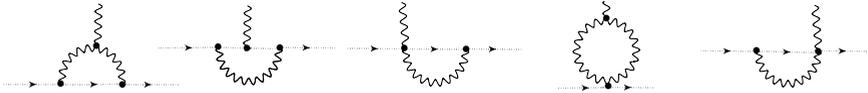}
\caption{One-loop Feynman diagrams contributing to $\langle c^{\alpha}(x) \bar{c}^{\beta}(y) u_\mu^{\gamma}(z)\rangle$.  
}
\label{3pt}
\end{figure}

The one-loop Feynman diagrams (one-particle irreducible (1PI)) contributing to the quark propagator, $\langle \psi(x) \bar \psi(y) \rangle$,  are shown in Fig.~\ref{quark2pt}, those contributing to the squark propagator, $\langle A_+(x) A^{\dagger}_+(y) \rangle$, in Fig.~\ref{squark2pt}. Identical results are obtained for $\langle A_+(x) A_+^{\dagger}(y) \rangle$ and $\langle  A_-^{\dagger}(x) A_-(y)\rangle$. The last diagram in Fig.~\ref{squark2pt} is responsible for mixing between $A_+$ and $A_-$. The one-loop Feynman diagrams contributing to the gluon propagator, $\langle  u_\mu^{\alpha}(x) u_\nu^{\beta}(y) \rangle$, and gluino propagator, $\langle  \lambda^{\alpha}(x) \bar \lambda^{\beta}(y) \rangle$, are shown in Fig.~\ref{gluon2pt} and Fig.~\ref{gluino2pt}, respectively. Lastly, the 1PI Feynman diagram which contributes to the ghost propagator, $\langle c(x) \bar c(y) \rangle$, is shown in Fig.~\ref{ghost2pt}. In this work we also calculate the gluon-antighost-ghost Green's function in order to renormalize the coupling constant. In Fig.~\ref{3pt} we have drawn the corresponding lattice 1PI Feynman diagrams for the 3-pt function.

As is usually done, we will work in a mass-independent scheme, and thus all of our calculations, in the continuum as well as on the lattice, will be done at zero renormalized masses for all particles.


\subsection{$\MSbar$-Renormalized Green's Functions}
          
The first step in our perturbative procedure is to calculate the 2-pt and 3-pt Green's functions in the continuum, where we regularize the theory in $D$ Euclidean dimensions ($D=4-2\,\epsilon$)~\cite{Jack:1997sr}. From this computation we obtain the $\MSbar$-renormalized Green's functions by elimination of the pole part of the continuum bare Green's functions. The renormalized Green's functions are relevant for the ensuing calculation of the corresponding Green's functions using lattice regularization and $\MSbar$ renormalization. Depending on the prescription used to define $\gamma_5$ in $D$-dimensions, mixing between squarks may appear also in dimensional regularization. For the continnum calculations we adopt the t'Hooft-Veltman (HV) scheme~\cite{tHooft:1972tcz}. Other prescriptions are related among themselves via finite conversion factors.
   
Here we collect all $\MSbar$-renormalized results for the 2-pt and 3-pt Green's functions; the first result which we present, is the inverse quark renormalized propagator in momentum space:
\be
\langle \tilde \psi^R(q) \tilde{\bar{\psi}}^R(q') \rangle^{\MSbar}_{\rm{inv}} = (2\pi)^4 \delta(q-q') i \qslash \left[ 1 + \frac{g^2\,C_F}{16\,\pi^2} \left( 
4 + \alpha +  (2+\alpha) \log\left(\frac{\bar\mu^2}{q^2} \right) \right) \right].
\label{GF2quark}
\ee
where $C_F=(N_c^2-1)/(2\,N_c)$ is the quadratic Casimir operator in the fundamental representation, $q$ is the external momentum in the Feynman diagrams, and $\bar\mu$ is the $\MSbar$ renormalization scale. Note also that a Kronecker delta for color indices is understood in Eqs.~(\ref{GF2quark})-(\ref{GF2ghost}).

In matrix notation, our results for the renormalized 2-pt Green's functions with external squark legs are:
\bea
\label{MIXmatrix}
\hspace{-0.5cm}\langle \tilde A^R(q) \tilde{A}^{R\,\dagger}(q') \rangle_{\rm{inv}}^{\MSbar}&=& (2\pi)^4 \delta(q-q')\Bigg[ q^2 \begin{pmatrix} 1 & 0\\ 0 & 1 \end{pmatrix} + q^2 \frac{g^2\,C_F}{16\,\pi^2} \left( \frac{16}{3} +  (1+\alpha) \log\left(\frac{\bar\mu^2}{q^2}\right)\right)\begin{pmatrix} 1 & 0 \\ 0 & 1 \end{pmatrix}\nonumber\\
&& \hspace{2cm}+ q^2 \frac{g^2\,C_F}{16\,\pi^2}\frac{4}{3}\begin{pmatrix} 0 & 1\\ 1 & 0 \end{pmatrix}\Bigg],
\eea
where $A^R$ is a 2-component column which contains the renormalized squark fields: $A^R = \left( {\begin{array}{c}  A^{R}_+ \\ {A_-^{R}}^{\dagger} \end{array} } \right)$.

We now turn to the gluon renormalized propagator. The contributions from the diagrams of Fig.~\ref{gluon2pt}, taken separately, are not transverse. But, their sum has this property, and it is found to take the following form:
\bea
\langle  \tilde u_\mu^{R}(q) \tilde u_\nu^{R}(q') \rangle^{\MSbar}_{\rm{inv}} &=&(2\pi)^4 \delta(q+q') \Bigg\{  \frac{1}{\alpha} q_{\mu} q_{\nu} + \left(q^2 \delta_{\mu \nu} - q_{\mu} q_{\nu}\right)\Bigg[ 1  + \frac{g^2\, N_f}{16\,\pi^2} \left( 
2 +\log\left(\frac{\bar\mu^2}{q^2} \right)\right)\nonumber\\
&& \hspace{2cm}- \frac{g^2\, N_c}{16\,\pi^2}\frac{1}{2} \left(
\frac{19}{6} + \alpha + \frac{\alpha^2}{2} + \left(3-\alpha\right)\log\left(\frac{\bar\mu^2}{q^2} \right)\right)\Bigg] \Bigg\}.
\eea
The result for the inverse gluino renormalized propagator to one-loop order is:
\be
\langle \tilde \lambda^{R}(q) \tilde{\bar{\lambda}}^{R}(q') \rangle^{\MSbar}_{\rm{inv}} = (2\pi)^4 \delta(q-q') \frac{i}{2}\, \qslash \Bigg[ 1 + \frac{g^2\,N_f}{16\,\pi^2}\left( 2 
+ \log\left(\frac{\bar\mu^2}{q^2} \right)\right)+ \frac{g^2\, N_c}{16\,\pi^2} \left(\alpha 
+ \alpha \log\left(\frac{\bar\mu^2}{q^2}  \right)\right)\Bigg].
\ee
The ghost propagator is the same as in the non-supersymmetric case:
\be
\langle \tilde c^{R}(q) \tilde{\bar{c}}^{R}(q') \rangle^{\MSbar}_{\rm{inv}} = (2\pi)^4 \delta(q-q') q^2 \left[ 1 - \frac{g^2\,N_c}{16\,\pi^2} \left(1 + 
\frac{1}{4} (3-\alpha) \log\left(\frac{\bar\mu^2}{q^2} \right) \right) \right].
\label{GF2ghost}
\ee
Lastly, the 3-pt amputated Green's function, at zero antighost momentum, in $\MSbar$ renormalization scheme, gives:
\be
\langle \tilde c^{R\,\alpha}(q) \tilde{\bar{c}}^{R\,\beta}(0) \tilde{u}_\mu^{R\,\gamma}(q')\rangle^{\MSbar}_{\rm{amp}}  = 
(2\pi)^4 \delta(q+q')\, f^{\alpha\,\beta\,\gamma} (i g q_\mu) \Bigg[1 + \frac{g^2\,N_c}{16\,\pi^2}\frac{\alpha}{2} \Bigg(1 
+  \log \left(\frac{\bar{\mu}^2}{q^2} \right)\Bigg)\Bigg].
\label{GF3ghostsgluon}
\ee

\subsection{Green's Functions on the lattice}

The first result presented here, Eq.~(\ref{GF2quarklatt}), is the lattice inverse quark propagator up to one loop. In all lattice expressions the systematic errors, coming from numerical loop integration, are smaller than the last digit we present.
\bea
\langle \tilde \psi^B(q) \tilde{\bar{\psi}}^B(q') \rangle^{L}_{\rm{inv}} &=& (2\pi)^4 \delta(q-q') \Bigg\{ i \qslash \left[1  - \frac{g^2\,C_F}{16\,\pi^2} \left[ 12.8025 - 4.7920 \alpha + (2+\alpha)\log\left(a^2\,q^2\right)\right]\right]\nonumber\\
&&\hspace{2cm} + \frac{g^2\,C_F}{16\,\pi^2}\frac{1}{a} 51.4347 \,r \Bigg\}.
\label{GF2quarklatt}
\eea              
The inverse squark propagator, is:
\bea
\langle  A^B A^{B\,\dagger}\rangle^{L}_{\rm{inv}} &=& q^2 \openone + \frac{g^2\,C_F}{16\,\pi^2}\Bigg\{ q^2\left[11.0173 - 3.7920 \alpha + (1+\alpha)\log(a^2\,q^2) \right] \begin{pmatrix} 1 & 0\\ 0 & 1 \end{pmatrix}\nonumber\\
&&\hspace{2cm} +1.0087 q^2 \begin{pmatrix} 0 & 1\\ 1 & 0 \end{pmatrix} + \frac{1}{a^2} \begin{pmatrix}65.3930 & 75.4031 \\ 75.4031 & 65.3930\end{pmatrix} \Bigg\}.
\label{GF2squarklatt}
\eea
The gluino inverse propagator is:
\bea
\label{GF2gluinolatt}
\langle \tilde \lambda^{B}(q) \tilde{\bar{\lambda}}^{B}(q') \rangle^{L}_{\rm{inv}} &=& (2\pi)^4 \delta(q-q') \Bigg\{ \frac{i}{2} \, \qslash \Bigg[ 1 + \frac{g^2\,N_f}{16\,\pi^2}\left(1.9209 - \log\left(a^2\,q^2\right)\right) \nonumber\\
&& \hspace{3cm} - \frac{g^2\, N_c}{16\,\pi^2} \left( 16.6444 - 4.7920\,\alpha  + \alpha \log\left(a^2\, q^2 \right)\right)\Bigg] \nonumber\\
&&\hspace{2cm}+ \frac{g^2}{16\,\pi^2} \frac{N_c}{2} \frac{1}{a} 51.4347 \,r\Bigg\}.
\eea
The gluon inverse propagator is given by:
\bea
\langle  \tilde u_\mu^{B}(q) \tilde u_\nu^{B}(q') \rangle^{L}_{\rm{inv}} &=&(2\pi)^4 \delta(q+q') \Bigg\{ \frac{1}{\alpha}  q_{\mu} q_{\nu} + \left(q^2 \delta_{\mu \nu} - q_{\mu} q_{\nu}\right)\Bigg[1 \nonumber\\
&&-\frac{g^2}{16\,\pi^2}\Bigg[ -19.7392 \frac{1}{N_c} + N_f\left(-2.9622 + \log\left(a^2\,q^2\right)\right)\nonumber\\
&& - N_c\left(20.1472 - 0.8863\,\alpha +  \frac{\alpha^2}{4} + \left(\frac{\alpha}{2} -\frac{3}{2}\right)  \log\left(a^2\, q^2 \right)\right)\Bigg]\Bigg]\Bigg\}.
\label{GF2gluonlatt}
\eea
The ghost field renormalization, $Z_c$, which enters the evaluation of $Z_g$  can be extracted from the ghost propagator:
\be
\langle \tilde c^{B}(q) \tilde{\bar{c}}^{B}(q') \rangle^{L}_{\rm{inv}}  =  (2\pi)^4 \delta(q-q') q^2 \left[ 1+\frac{g^2\,N_c}{16\,\pi^2}\left(4.6086 - 1.2029 \alpha -\frac{1}{4}\left( 3 - \alpha \right)
\log\left(a^2\,q^2\right)\right)\right].
\label{GF2ghostlatt}
\ee
Lastly, the amputated gluon-antighost-ghost Green's function is:
\be
\langle \tilde c^{B\,\alpha}(q) \tilde{\bar{c}}^{B\,\beta}(0) \tilde{u}_\mu^{B\,\gamma}(q')\rangle^{L}_{\rm{amp}}  = 
(2\pi)^4 \delta(q+q')\, f^{\alpha\,\beta\,\gamma} \left(i g  q_\mu\right) \Big[1 + \frac{g^2\,N_c}{16\,\pi^2}\left(
2.3960 \alpha - \frac{1}{2} \alpha \log\left(a^2\, q^2 \right)\right)
\Big].
\label{GF3ghostsgluon}
\ee
The critical masses for the quark, squark and gluino can be read off (up to a minus sign) from the ${\cal O}(q^0)$ parts of Eqs. (\ref{GF2quarklatt}), (\ref{GF2squarklatt}), (\ref{GF2gluinolatt}), respectively. 

\subsection{Renormalizarion Factors}
Renormalization factors relate bare quantities ($B$) on the lattice to their renormalized ($R$) continuum counterparts: 
\bea
\hspace{5cm} \psi^R &=& \sqrt{Z_\psi}\,\psi^B,\\
A^R_\pm &=& \sqrt{Z_{A_\pm}}\,A^B_\pm, \\
u_{\mu}^R &=& \sqrt{Z_u}\,u^B_{\mu},\\
\la^R &=& \sqrt{Z_\la}\,\la^B,\\
c^R &=& \sqrt{Z_c}\,c^B, \\
g^R &=& Z_g\,\mu^{-\epsilon}\,g^B.
\eea
Combining the Green's functions on the lattice with the corresponding results from the continuum, we extract $Z_\psi^{L,\MSbar}$, $Z_u^{L,\MSbar}$, $Z_\lambda^{L,\MSbar}$, $Z_{A_\pm}^{L,\MSbar}$, $Z_c^{L,\MSbar}$ and $Z_g^{L,\MSbar}$ in the $\MSbar$ scheme and on the lattice. 
\\               
\be
Z_\psi^{L,\MSbar} = 1 + \frac{g^2\,C_F}{16\,\pi^2} \left( -16.8025 + 3.7920 \alpha - (2+\alpha)\log\left(a^2\,\bar\mu^2\right) \right).
\ee

\be
Z_{\lambda}^{L,\MSbar} =  1 - \frac{g^2\,}{16\,\pi^2} \left[N_c\left(16.6444 - 3.7920 \alpha +  \alpha \log\left(a^2\,\bar\mu^2\right)\right)+ N_f\left(0.07907 + \log\left(a^2\,\bar\mu^2\right)\right) \right].
\ee

\be
\left(Z_A^{1/2}\right)^{L,\MSbar}= \openone - \,\frac{g^2\,C_F}{16\,\pi^2}\Bigg\{\Bigg[8.1753 - 1.8960\alpha+\frac{1}{2}(1+\alpha)\log\left(a^2\,\bar\mu^2\right)\Bigg] \begin{pmatrix} 1 & 0\\ 0 & 1 \end{pmatrix} - 0.1623 \begin{pmatrix} 0 & 1\\ 1 & 0 \end{pmatrix}\Bigg\}.
\ee

\bea
Z_{u}^{L,\MSbar} &=&  1 + \frac{g^2\,}{16\,\pi^2} \Big[19.7392\frac{1}{N_c}- N_c\left(18.5638 - 1.3863  \alpha + \left( -\frac{3}{2}+\frac{\alpha}{2}\right) \log\left(a^2\,\bar\mu^2\right)\right)\nonumber\\
&&\hspace{1.3cm}+ N_f\left(0.9622   - \log\left(a^2\,\bar\mu^2\right)\right)\Big].
\eea

\be
Z_c^{L,\MSbar} = 1 - \frac{g^2 N_c}{16\pi^2} \Bigl[3.6086 - 1.2029 \alpha -\frac{1}{4}\left( 3 - \alpha \right) \log\left(a^2\,\bar{\mu}^2\right) \Bigr].
\ee

\be
Z_g^{L,\MSbar} =  1 + \gtilde\,\Bigl[ -9.8696 \frac{1}{N_c} + N_c \left( 12.8904  - \frac{3}{2} \log\left(a^2\,\bar{\mu}^2\right)\right)-\,N_f\left( 0.4811 - \frac{1}{2} \log(a^2\,\bar{\mu}^2)\right)\Bigr].
\ee
From the calculation of $Z_{g}^{L,\MSbar}$ one can extract the Callan-Symanzik beta-function for SQCD. On the lattice, the bare beta-function is defined as: 
$\beta_L(g^B)=-a dg^B/da|_{g^R,\,\bar{\mu}}$. 
The first term in this expansion is:
\be
\beta_L(g)=\frac{g^{3}\,}{16\,\pi^2} \left(-3N_c + N_f\right)+{\cal{O}}(g^5).
\ee
For $N_f < 3N_c$, the ${\cal{O}}(g^3)$ term is negative, in other words, the theory is asymptotically free. Our finding for the beta function agrees with what is obtained in the supersymmetric Yang-Mills theory~\cite{DRT}.
 
\section{Future Plans -- Conclusion}
\label{future}

In this work we have performed a pilot investigation of issues related to the formulation of a supersymmetric theory on the lattice. As a prototype model, we have  studied ${\cal N} = 1$ supersymmetric QCD. This model bears all major characteristics of potential extensions of the standard model, including superpartners for gauge and matter fields; it is thus appropriate for a feasibility study on the lattice.

There are several well-known problems arising from the complete (or even partial) breaking of supersymmetry in a regularized theory, including the necessity for fine tuning of the theory's bare Lagrangian, and a rich mixing pattern of composite operators at the quantum level. We address these problems via perturbative calculations at one loop. In order to provide the necessary ingredients for performing numerical studies of supersymmetric theories, we have calculated the self energies of all particles which appear in SQCD. We determined the renormalization factors for these fields; in addition, for the squark propagator we found the mixing coefficients among its different degrees of freedom. Furthermore, we have computed the gluon-antighost-ghost Green's function in order to renormalize the coupling constant. Our results are also relevant to the investigation of relationships between different Green's functions involved in SUSY Ward identities \cite{Taniguchi:2000, Vladikas:2002}.
   
There are several directions in which this work could be extended. A natural extension would be the computation of the Green's functions for composite operators made of quark, squark, gluon and gluino fields; studies of such operators in the continuum can be found in, e.g., Refs.~\cite{Konishi, Rattazzi, Silvestroni, Matthias}. A serious complication in the supersymmetric case regards the mixing of quark bilinear operators with other composite operators. A whole host of operators with equal or lower dimensionality, having the same quantum numbers and same transformation properties can mix at the quantum level; on the lattice, the number of operators which mix among themselves is considerably greater than in the continuum regularization. We are planning to study their renormalization and mixing perturbatively. The perturbative computation of all relevant Green's functions of these operators, will be followed by the construction of the mixing matrix, which may also involve nongauge invariant (but BRST invariant) operators or operators which vanish by the equations of motion. 

Finally, it would be important to extend our computations to further improved actions with reduced lattice artifacts and reduced symmetry breaking, e.g. the overlap fermion action, as a forerunner to numerical studies using these actions.

\bibliography{lattice2017proc}

\end{document}